# Assessing the Influence of *d*-Orbital Radius on Formation of Localized Photogenerated States in Corundum Metal Oxides

**Erica P. Craddock and Kathryn E. Knowles***

*Department of Chemistry, University of Rochester, Rochester NY 14627, USA*

*corresponding author e-mail: kknowles@ur.rochester.edu

## ABSTRACT

Photogenerated polarons are fundamental to the photophysics of transition metal oxide semiconductors. It is therefore imperative to understand the mechanisms by which polarons form upon photoexcitation of transition metal oxides to realize their potential in photoapplications. Hematite ($\alpha$-$Fe_2O_3$) is known to form photoexcited small polarons, which limit its performance as a photoelectrocatalyst for water oxidation. Here, we report a systematic comparison of the electronic, optical and vibrational properties of hematite to those other metal oxides in the corundum crystal family that elucidates the impact of *d*-orbital radius on carrier-phonon coupling. Three corundum metal oxides are analyzed: $\alpha$-$Al_2O_3$ (no *d*-electrons), $\alpha$-$Fe_2O_3$ (3*d*), and $\alpha$-$Rh_2O_3$ (4*d*) with a combined approach of resonance Raman spectroscopy, thermal difference optical spectroscopy, and computational modeling of electronic and vibrational states. We find that the Raman spectrum of $\alpha$-$Al_2O_3$ does not change as the Raman excitation is varied across the visible region, as there is no optical absorption. In contrast, both $\alpha$-$Fe_2O_3$ and $\alpha$-$Rh_2O_3$ exhibit strong coupling of phonons to optical transitions at the onset of absorption, which is evidence of excitation into a polaronic state. Closely comparing the optical polaronic properties of $\alpha$-$Fe_2O_3$ and $\alpha$-$Rh_2O_3$, we establish that increased lattice covalency in $\alpha$-$Rh_2O_3$ arising from the increased radial extension of the 4*d* orbitals influences which phonon modes mediate photogenerated polaron formation.

## INTRODUCTION

Transition metal oxide semiconductors are widely studied for photoapplications such as photo(electro)catalysis[1,2] and optoelectronics.[3] Semiconducting metal oxides containing first-row transition metals are particularly attractive due to the earth-abundance of their raw materials, low to non-toxic compositions, redox-active cations and optical band gaps lying within the visible range.[1,4] However, the weakly dispersive nature of 3*d* bands and strong electron-phonon coupling in first-row transition metal oxides makes them prone to forming small polarons, which significantly impacts their photophysics.[5–7] A polaron is a quasiparticle that forms upon the coupling of a carrier (electron, hole) with lattice vibrations (phonons) creating a distorted lattice and an energetically and spatially isolated electronic state.[8–10] In the case of hematite ($\alpha$-$Fe_2O_3$), the formation of photogenerated small polarons limits its ability to be an efficient photoelectrocatalyst for water oxidation.[11–14] Surface polaronic states in $TiO_2$ nanoparticles have been shown to modulate binding of small molecules to the nanocrystal surface,[15–17] which has potential implications for catalytic performance. In general, due to the recent recognition that transition metal oxides tend to form polaronic states upon photoexcitation, some progress has been made toward explicitly utilizing photogenerated polaronic states in transition metal oxides for photocatalysis.[18,19]

There are two predominant modern polaron models that extend the original Landau-Pekar polaron theory[8,9] into a quantum mechanical framework: the Frölich model, which describes large polarons using long-range carrier-phonon coupling following a continuum approximation;[20,21] and the Holstein model, which accounts for the discreteness of the lattice and describes short-range carrier-phonon coupling to form small polarons.[22,23] A small (large) polaron is defined as one in which the volume of the lattice deformation is smaller (larger) than the volume of a unit cell of the material. Temperature-dependent conductivity studies in conjunction with the Frölich and Holstein models have shaped the current understanding of polaron transport mechanisms;[10,24] however, in both theoretical and experimental investigations of polaron transport, only a single carrier type (i.e. electron or hole) is introduced to the system. With photoexcited carriers, by definition, both an electron and a hole are generated, enabling multiple carrier types in the system to form polaronic states simultaneously, a scenario that was not considered when the Frölich and Holstein models were developed. Furthermore, recent reports using density functional theory (DFT) examine the formation of "medium-sized" polarons in $WO_3$, which does not fit the traditional models developed by Frölich and Holstein, and is instead described as an "anti-distortive" polaron due to the local "undoing" of symmetry-lowering distortions by the polaronic lattice displacements.[25] Clearly, there is much left to be understood about the factors that determine polaron formation and properties within photoexcited metal oxide systems.

Spectroscopic and computational reports indicate that the weakly dispersive 3*d* bands in $\alpha$-$Fe_2O_3$ and other first-row transition metal oxides such as $Co_3O_4$ and $BiVO_4$, lead to formation of highly localized small (Holstein) polarons upon photoexcitation.[6,7,26–31] The Fe-O bonds in $\alpha$-$Fe_2O_3$ are ionic in nature and, consequently, photogenerated electrons (holes) become localized to Fe (O) bands. We hypothesize that the degree of orbital overlap in a lattice (the lattice covalency) significantly impacts the formation of photogenerated polarons. Here, we aim to test this hypothesis by comparing the photophysical properties of three binary metal oxides that all have the same crystal structure (corundum, Figure 1A) but vary in the involvement of *d*-orbitals in the band-edge states: $\alpha$-$Al_2O_3$, $\alpha$-$Fe_2O_3$ and $\alpha$-$Rh_2O_3$. $Al^{3+}$ has no *d* electrons whereas $Fe^{3+}$ has a partially filled 3*d* shell and $Rh^{3+}$ has a partially filled 4*d* shell. Figure 1B plots the radial distribution function of 3*d* and 4*d* wavefunctions, which describe the valence orbitals of the $Fe^{3+}$ and $Rh^{3+}$ ions in $\alpha$-$Fe_2O_3$ and $\alpha$-$Rh_2O_3$, respectively. In general, 4*d* orbitals are larger and more diffuse than 3*d* orbitals, which provides the opportunity for greater overlap with oxygen 2*p* orbitals. Using a combined approach of DFT, Stokes resonance Raman spectroscopy, and thermal difference optical spectroscopy, we demonstrate that the phonon modes that mediate the localization of photogenerated states are directly influenced by lattice covalency.

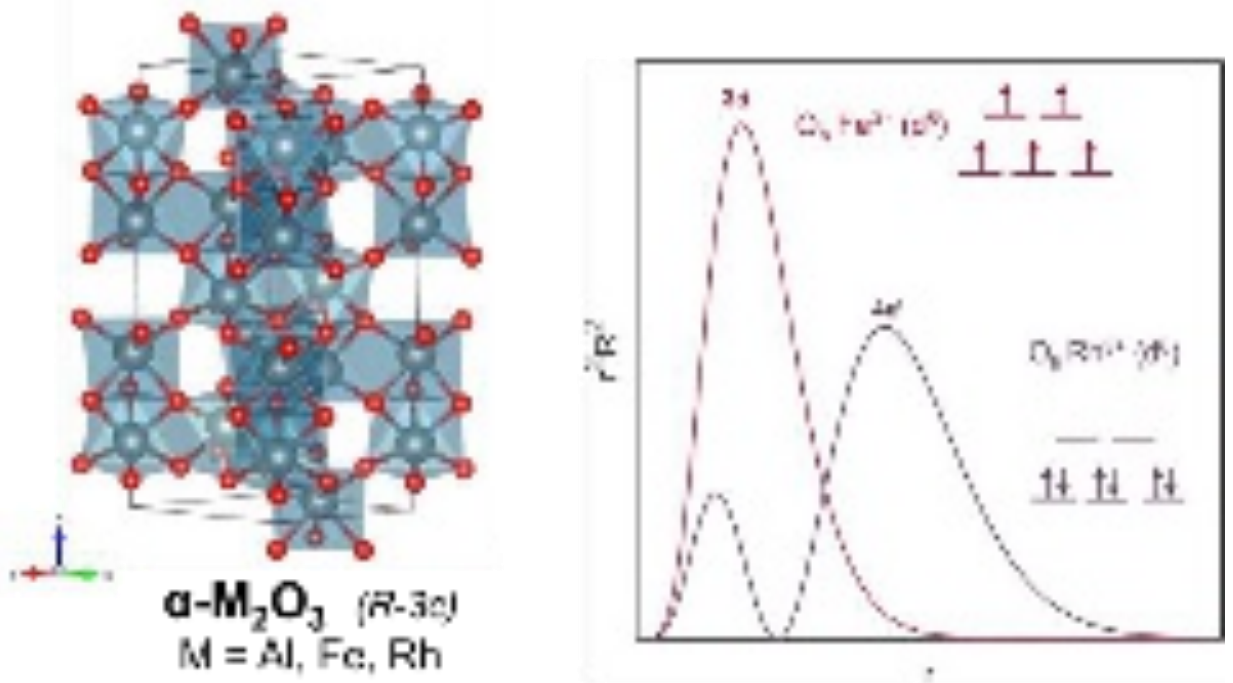

**Figure 1. (A)** Crystal structure of corundum group metal oxides with $R\bar{3}c$ symmetry. **B)** Radial distribution functions of 3*d* and 4*d* orbitals, which constitute the valence shells of the $Fe^{3+}$ and $Rh^{3+}$ ions in $\alpha$-$Fe_2O_3$ and $\alpha$-$Rh_2O_3$. The electron configurations of these ions within the octahedral crystal field of the corundum lattice are shown in the inset.

## EXPERIMENTAL METHODS

*Density Functional Theory:*

The electronic structures of the corundum oxide materials were computed using density functional theory (DFT), performed using the pseudopotential plane wave package implemented in Quantum ESPRESSO.[51–53] Depending on the metal in the crystal structure, Hubbard (*U*) and Hund (*J*) corrections were calculated using a linear response method,[7,33] then applied to the electronic calculations. The Hubbard *U* value corrects for electron correlation[33] while the Hund *J* value corrects for magnetic interactions.[34] The Fe cations in α-$Fe_2O_3$ have a high-spin $d^5$ electron configuration (Fig. 1) and therefore both Hubbard and Hund parameters are applied to Fe. Conversely, the Rh cations in α-$Rh_2O_3$ have a quasi-closed, low-spin $d^6$ electron configuration with no unpaired electrons (Fig. 1) and therefore only a Hubbard parameter was applied to Rh. With no *d* electrons present in α-$Al_2O_3$, no Hubbard or Hund parameters were applied in its calculation. Table 1 contains the values of the Hubbard and Hund parameters used in our calculations. Specifications for all α-$Fe_2O_3$ calculations can be found in previous reports.[6,7] High plane wave cut-off energies (≤ 1100 eV) were used to ensure the total energy and total interatomic forces were converged to 10 meV and 0.10 meV/Å. α-$Al_2O_3$ and α-$Rh_2O_3$ were both sampled over 5 x 5 x 5 k-point grids. Both the Perdew, Burke, and Ernzerh exchange-correlation functional revised for solids (PBEsol)[54,55] and Optimized Norm-Conserving Vanderbilt (ONCV) [56,57] method for pseudopotentials were implemented throughout this work. Iterative geometric relaxations were performed using the Broyden–Fletcher–Goldfarb–Shanno (BFGS) algorithm.[58–60] Electronic bands and projected densities of states were then determined for the relaxed structures by first calculating the self-consistent field (SCF) wavefunctions with associated Hubbard and Hund corrections and then performing a Fourier interpolation to a larger grid of k-points. Phonon modes of α-$Al_2O_3$ and α-$Rh_2O_3$ were calculated with density functional perturbation theory (DFPT) using the PHonon code implemented in Quantum ESPRESSO.[51–53] Starting with the same relaxed ground-state configuration used to calculate the electronic states, dynamical matrices were evaluated at the high-symmetry point Γ. In the case of α-$Rh_2O_3$, the vibrational density of states was determined from evaluating a uniform 3 × 3 × 3 grid of q-points followed by finer sampling via Fourier interpolation on a denser q-grid.

**Table 1.** Hubbard and Hund Corrections Applied to α-$Fe_2O_3$ and α-$Rh_2O_3$

| **α-$Fe_2O_3$*** | **α-$Rh_2O_3$** |
|---|---|
| **U:** 3.12045305844106<br>**J:** 1.53503152150523 | **U:** 4.30741117738007 |

*Reported in references [6,7].

*Sample Preparation*

α-$Rh_2O_3$ and α-$Fe_2O_3$ thin films were prepared using 3 mmol of nitrate salt ($Rh(NO_3)_3$ hydrate, $Fe(NO_3)_3$ hexahydrate) combined with 4.5 mmol citric acid monohydrate and 4.5 mmol ethylene glycol in 6 mL of ethanol. Aliquots (50 mL) of precursor solutions were spin-coated at 5000 rpm onto 1-mm thick substrates (sapphire and fused quartz) and annealed at 600 °C for 5 minutes. Multiple layers were deposited in this manner before a final annealing at 600 °C for 24 hours followed by slow cooling to room temperature in the furnace. One annealed aliquot contributes ~40 nm to film thickness. Sapphire substrates were used for all α-$Al_2O_3$ data presented and were purchased from Esco optics (G107040, thickness = 1 mm).

*Optical Spectroscopy*

Transmission and reflection spectra were collected from 2460 – 200 nm with an Agilent Cary 7000 Spectrometer equipped with a Universal Measurement Accessory. The samples were set to a 10° angle relative to the incident beam and transmission spectra were collected with the detector at 180° while reflection spectra were collected with a 20° detector angle. Temperature-dependent spectra above room temperature were collected using a home-built setup in which the sample is sandwiched between two donut-shaped ceramic heaters, which allow optical access through their centers. The temperature was controlled using an Omega temperature controller. For spectra collected below room temperature, an APD Cryogenics Inc. Model LT-3-110 Heli-Tran cryostat system run at $10^{-5}$ Torr vacuum with a turbo molecular vacuum pump was used. The cryostat sleeve is equipped with a quartz window for transmission collection and a large sapphire window to accommodate small-angle reflection collection. Using a previously reported Fresnel analysis method,[6,32] the imaginary dielectric spectra of the samples were calculated for all temperatures in which transmission and reflection spectra were measured.

*Raman spectroscopy*

Resonance Raman spectra were collected using CW diode lasers with wavelengths of 660 nm (1.88 eV), 594 nm (2.09 eV), 561 nm (2.21 eV), 532 nm (2.33 eV), 515 nm (2.41 eV) and 491 nm (2.53 eV) and Princeton Instruments monochromators equipped with a CCD detector. Laser power was kept $< 30$ mW to limit the possibility of sample heating. The Raman shift axis was corrected with cyclohexane spectra. For Figure 6, the Raman intensity of both α-$Rh_2O_3$ and α-$Fe_2O_3$ spectra collected at each excitation wavelength was corrected for grating efficiency and scattering cross section.

## RESULTS AND DISCUSSION

### Optical Properties and Electronic Structure of Corundum Metal Oxides

Polycrystalline thin films of α-$Fe_2O_3$ and α-$Rh_2O_3$ were fabricated on commercially available sapphire (α-$Al_2O_3$) substrates using previously reported sol-gel techniques (see Methods). A bare substrate was used as the α-$Al_2O_3$ sample. Figure 2 plots the experimental imaginary dielectric spectra of α-$Al_2O_3$, α-$Fe_2O_3$ and α-$Rh_2O_3$ extracted by applying the Fresnel equations to a series of transmission and reflection spectra.[6,32] With no *d*-electrons, α-$Al_2O_3$ effectively has no electronic transitions in the visible range at energies below 4 eV, which is further confirmed by the colorless appearance of the film (Fig. 2, inset). Although both α-$Fe_2O_3$ and α-$Rh_2O_3$ exhibit absorption throughout the visible region, the absorption onset occurs at a lower energy for α-$Rh_2O_3$ and it also has a lower intensity throughout the visible range compared to α-$Fe_2O_3$. As discussed in detail below, these differences are a direct consequence of the different *d*-electron configurations in these materials.

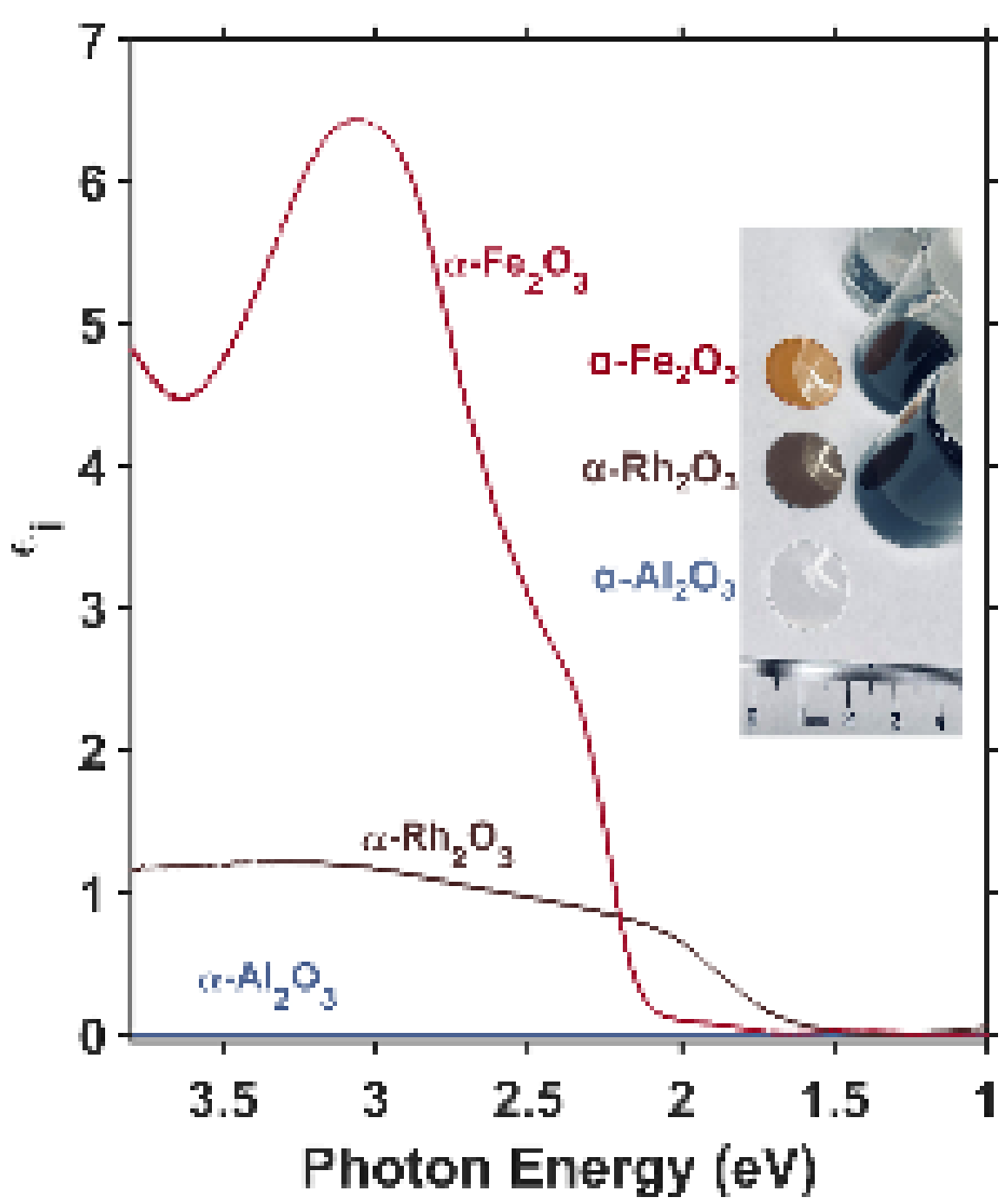

**Figure 2. A)** Imaginary dielectric spectrum of α-$Al_2O_3$ (blue), α-$Fe_2O_3$ (red) and α-$Rh_2O_3$ (maroon) determined experimentally from Fresnel analysis. *Inset*: Photos of corundum metal oxide thin films. The reflected image of the scintillation vials placed next to the films are captured on the surface of the films

demonstrating their specularly reflective properties. Imaginary dielectric spectrum of α-$Fe_2O_3$ in (**A**) is reproduced with permission from reference [6]. Copyright 2021 American Chemical Society.

The electronic structure and optical spectra of the three corundum metal oxides were calculated using DFT. Depending on the electronic configuration of the metal, different correction factors are applied in the DFT calculations of each corundum metal oxide. The Hubbard $U$ value corrects for electron correlation[33] while the Hund $J$ value corrects for magnetic interactions;[34] both correction factors are typically applied to transition metals.[35] The Fe cations in α-$Fe_2O_3$ have a high-spin $d^5$-electron configuration (Fig. 1B) and therefore both Hubbard and Hund parameters are applied to Fe. Conversely, the Rh cations in α-$Rh_2O_3$ have a quasi-closed, low-spin $d^6$-electron configuration with no unpaired electrons (Fig. 1B) and therefore only a Hubbard parameter was applied. With no occupied *d*-orbitals present in α-$Al_2O_3$, no Hubbard or Hund parameters were applied in its calculation. With *d*-electrons present in the crystal structure, weakly dispersive bands emerge at the edge of the conduction bands of both α-$Fe_2O_3$ and α-$Rh_2O_3$, which are not present in α-$Al_2O_3$ (Fig 3A-C). Both α-$Fe_2O_3$ and α-$Rh_2O_3$ exhibit dispersive bands of primarily *s*- and *p*-character that are also present in α-$Al_2O_3$ at higher energies in the conduction band; however, the presence of partially filled 3*d* and 4*d* orbitals contribute to weakly dispersive bands dominating the conduction band-edge. The 4*d* bands in α-$Rh_2O_3$ have overall higher dispersion than the 3*d* bands in α-$Fe_2O_3$, however these conduction bands are still relatively flat indicating that conduction band electrons in α-$Rh_2O_3$ and α-$Fe_2O_3$ have higher effective masses than those in α-$Al_2O_3$.[36] On the energy axis, the Rh 4*d* bands are less densely spaced, leading to a lower localized projected density of Rh character compared to Fe at the conduction band-edge. Additionally, there is significant Rh 4*d* and O 2*p* character throughout both the valence and conduction bands indicating a higher degree of covalency in α-$Rh_2O_3$ compared to both α-$Al_2O_3$ and α-$Fe_2O_3$.

**Figure 3.** Electronic bands and projected densities of states of **(A)** α-$Al_2O_3$, **(B)** α-$Fe_2O_3$, and **(C)** α-$Rh_2O_3$. The imaginary dielectric spectrum calculated with DFT overlaid with experimentally determined spectrum of **(C)** α-$Al_2O_3$, **(D)** α-$Fe_2O_3$ and **(E)** α-$Rh_2O_3$. Plots (**B**) and (**E**) reconfigured with permission from reference [6]. Copyright 2021 American Chemical Society.

The DFT-computed imaginary dielectric spectra of α-$Al_2O_3$, α-$Fe_2O_3$ and α-$Rh_2O_3$ are shown in Figures 3D-F overlaid with experimental spectra. The match of experimental and computational spectra of both α-$Al_2O_3$ and α-$Rh_2O_3$ demonstrate the accuracy of the electronic structure. In the case of α-$Fe_2O_3$, the shoulder at ~2.2 eV observed experimentally is not captured by the spectrum computed with DFT+*U*+*J*. Previous work from our group demonstrates that, upon incorporation of a stochastic average of room-temperature vibrational contributions to the electronic structure of α-$Fe_2O_3$, the shoulder at ~2.2 eV observed experimentally is captured computationally.[7] The electronic structures and optical spectra shown in Figure 3 demonstrate that the optical band gap decreases with increased principal quantum number of the *d* orbitals: α-$Al_2O_3$ has the widest band gap at 6.0 eV, the computed band gap of α-$Fe_2O_3$ is 2.4 eV, and α-$Rh_2O_3$ has a value of 1.3 eV. The experimental dielectric intensity of the band-gap transition also trends with *d*-shell character: α-$Fe_2O_3$ has the highest projected density of electronic character at the band edge, yielding the most intense optical transition, which is assigned as a ligand to metal charge transfer (LMCT)-type transition from a valence band that is primarily O 2*p* in character to a conduction band-edge that is primarily Fe 3*d* in character (Fig. 3). Both α-$Al_2O_3$ and α-$Rh_2O_3$ have lower partial density of states at the conduction band-edge, and it follows that the oscillator strength of the optical transition is lower as well. The primary optical transition in α-$Al_2O_3$ is also an LMCT-type transition from a valence band dominated by O 2*p* character to a conduction band dominated by Al 3*s* and 3*p* character. In α-$Rh_2O_3$, both the conduction and the valence bands have significant contributions from both the O 2*p* and the Rh 4*d* orbitals, which is indicative of a higher degree of lattice covalency than either α-$Al_2O_3$ or α-$Fe_2O_3$ (*vide infra*). The LMCT or metal-to-metal charge transfer (MMCT) designations used for optical transitions in more ionic first-row transition metal oxides[28,37–39] thus do not offer an accurate description of the optical transitions in α-$Rh_2O_3$.

**Optical Phonon Modes of Corundum Metal Oxides**

The Raman spectra of polycrystalline thin films of α-$Al_2O_3$, α-$Fe_2O_3$ and α-$Rh_2O_3$ are shown in Figure 4A. The primitive cell of these corundum metal oxides contains four metal cations and six oxygen anions, indicating there are a total of 30 phonon modes, of which 27 are optical phonon modes (Figure S1 in the supplementary material). Of the total optical phonons, there are seven Raman-active irreducible representations ($2A_{1g} + 5E_g$) for each corundum metal oxide; however, due to the varied atomic masses of the metals and different lattice bonding strengths, these phonons exist at different frequencies for each oxide material.[40–42] Using DFT-calculated phonon-displacement vectors at k-point Γ, the frequencies of the seven modes with Raman-active irreproducible representations were calculated for α-$Al_2O_3$ and α-$Fe_2O_3$ and matched to the experimentally observed Raman peaks. These assignments corroborate previous reports of polarization-dependent Raman spectra of single-crystal corundum metal oxides.[40–43] Irreproducible representations were assigned to the Raman peaks of α-$Rh_2O_3$ using the same method as α-$Al_2O_3$ and α-$Fe_2O_3$, however there are limited reports of α-$Rh_2O_3$ Raman spectra, none of which are of single-crystalline samples.[42] Given the accuracy of the phonon assignments with respect to the energetic ordering of the seven Raman-active irreducible representations derived from our DFT calculations of α-$Al_2O_3$ and α-$Fe_2O_3$, we trust the assignments of the phonon modes for α-$Rh_2O_3$ reported in Figure 4B, which are also determined from our DFT calculations. There is a weak feature observed at a Raman shift of 83 meV in the α-$Rh_2O_3$ spectrum that has been previously associated with an impurity of the orthorhombic phase.[42] With appreciable quantities of orthorhombic phase impurities, Weber *et al*. find many additional weak features contributing to the Raman spectrum of trigonal α-$Rh_2O_3$, which we do not observe.[42] The phase transition to orthorhombic occurs at 750 °C[44] and the α-$Rh_2O_3$ films in this work were annealed at 600 °C, decreasing the probability that impurity phases will form. There is a weak feature in the powder X-ray diffraction pattern of the α-$Rh_2O_3$ films likely associated with the orthorhombic phase, but all other peaks match the trigonal phase crystal reflections, demonstrating there are only minor contributions from the orthorhombic phase impurity (Figures S3 – S4 in the supplementary material) The DFT-computed phonon energies of all three corundum oxides are overlaid with the respective Raman spectra in Figure S2 in the supplementary material. It should be noted that the low-frequency phonon modes at 36.1 and 37.0 meV in α-$Fe_2O_3$ become resolved with changes incident polarization,[40,43,45] which was not assessed in this work due to the polycrystalline nature of our samples. However, the fundamental modes are still present and appear as a peak at 37.0 meV with a weak shoulder at 36.1 meV.

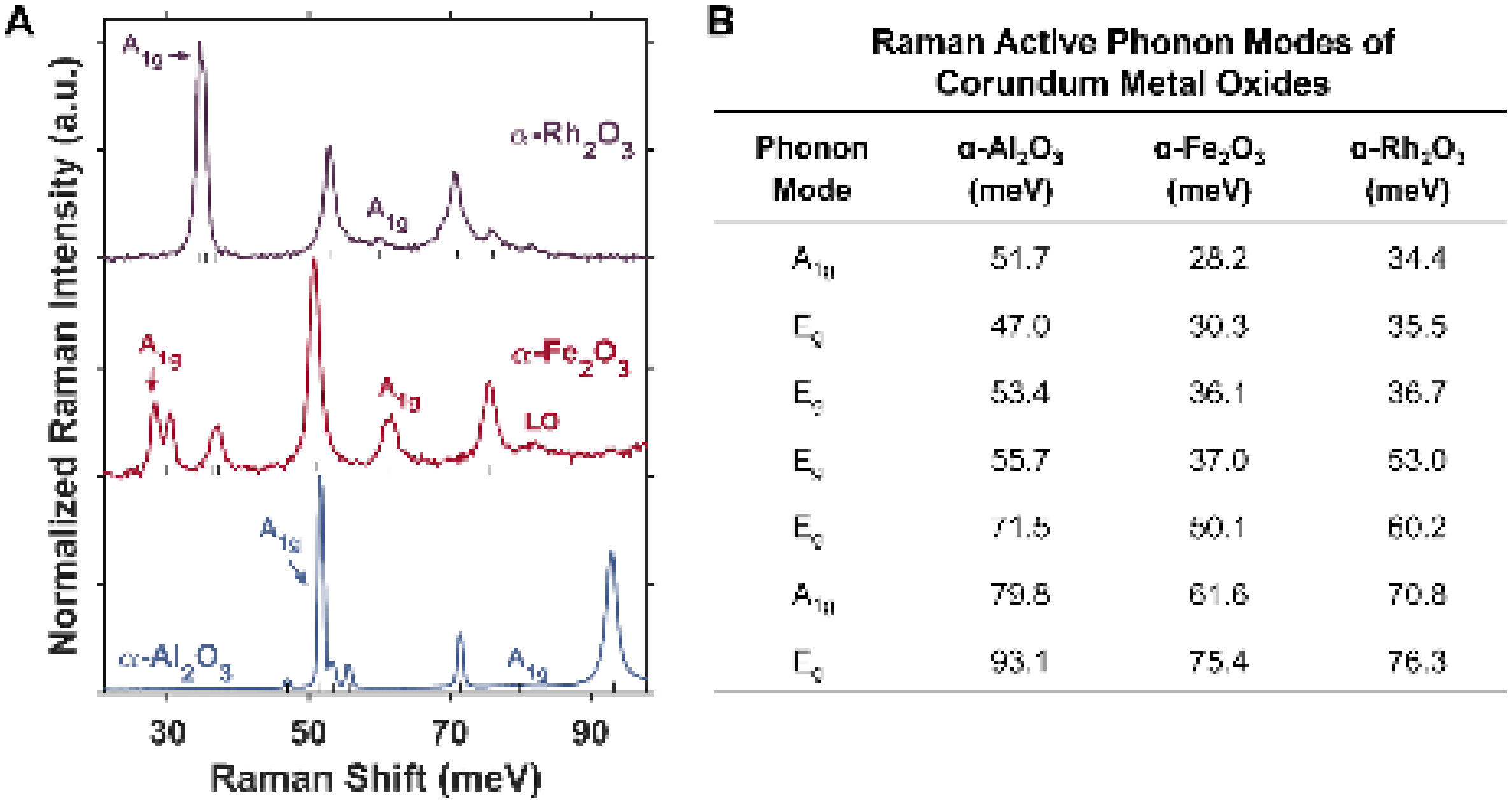

Raman Active Phonon Modes of Corundum Metal Oxides

| Phonon Mode | $\alpha$-$Al_2O_3$ (meV) | $\alpha$-$Fe_2O_3$ (meV) | $\alpha$-$Rh_2O_3$ (meV) |
|---|---|---|---|
| $A_{1g}$ | 51.7 | 28.2 | 34.4 |
| $E_g$ | 47.0 | 30.3 | 35.5 |
| $E_g$ | 53.4 | 36.1 | 36.7 |
| $E_g$ | 55.7 | 37.0 | 53.0 |
| $E_g$ | 71.5 | 50.1 | 60.2 |
| $A_{1g}$ | 79.8 | 61.6 | 70.8 |
| $E_g$ | 93.1 | 75.4 | 76.3 |

**Figure 4. A)** Stokes Raman spectra of $\alpha$-$Al_2O_3$ substrate (blue), $\alpha$-$Fe_2O_3$ polycrystalline thin film (red) and $\alpha$-$Rh_2O_3$ polycrystalline thin film (maroon). The ticks below the Raman spectra denote the peak centers of the distinct phonon modes of each material. The Raman spectra were collected with excitation energies of 2.71 eV ($\alpha$-$Al_2O_3$), 2.41eV ($\alpha$-$Fe_2O_3$), and 2.09 eV($\alpha$-$Rh_2O_3$). **B)** Table listing the frequencies and assignments of the Raman-active phonon modes of $\alpha$-$Al_2O_3$, $\alpha$-$Fe_2O_3$, and $\alpha$-$Rh_2O_3$.

Comparing the Raman spectra of the three corundum metal oxides, the energies of the phonon modes of $\alpha$-$Al_2O_3$ are higher than their counterparts in $\alpha$-$Fe_2O_3$ or $\alpha$-$Rh_2O_3$, likely due to the lighter atomic mass of Al. According to the phonon mode assignments of the corundum metal oxides, the lowest-energy phonon mode in $\alpha$-$Al_2O_3$ has $E_g$ symmetry, whereas the lowest energy modes in $\alpha$-$Fe_2O_3$ and $\alpha$-$Rh_2O_3$ have $A_{1g}$ symmetry. These assignments demonstrate that the atomic mass of the metal present in the structure and differences in lattice bonding strengths not only dictate the energies of the Raman-active phonon modes, but also the relative ordering of the modes in energy (Fig. 4). The seven modes of $\alpha$-$Al_2O_3$ are all energetically resolved from one another, even the less intense "shoulder" features at 53.4 and 55.7 meV closest to the most intense $A_{1g}$ peak. Contrastingly, the phonon modes with energies below 50 meV in $\alpha$-$Fe_2O_3$ from two closely spaced pairs, each with energy differences of less than 2 meV. In $\alpha$-$Rh_2O_3$, the lowest energy modes are even more closely spaced, with three different phonon modes contributing to the feature around 35 meV in $\alpha$-$Rh_2O_3$, with energetic spacing of about 1 meV. This trend in the energetic resolution of low-energy phonon modes in corundum metal oxides demonstrates that with a larger difference in atomic mass of metal and oxygen, the phonon modes become less distinguishable energetically. In metal oxides, low-frequency Raman modes are dominated by metal motion and with increasing Raman shift energy, the phonon displacements become dominated by oxygen motion.[46,47]

Figure 5 plots normalized Stokes Raman spectra of α-$Al_2O_3$ (A), α-$Fe_2O_3$ (B), and α-$Rh_2O_3$ (C) collected with a variety of excitation photon energies. The relative intensities of the α-$Al_2O_3$ phonon modes do not change significantly with different excitation energies. In contrast, the relative intensities of the phonon modes of α-$Fe_2O_3$ change drastically with different photon excitation energies: with 2.53-eV excitation, the 50-meV phonon is most prominent, but this mode decreases in intensity with decreasing excitation photon energy. The low-energy modes (28-meV and 30-meV) become more intense relative to the 50-meV mode with 2.21-eV excitation. Additionally, a prominent second order feature at 163 meV appears with 2.41-eV excitation and remains present until 2.09-eV excitation. This 163-meV mode corresponds to a second-order scattering of the longitudinal optical (LO) mode centered at 81 meV.[6] The resonance enhancement of this mode observed when excited with 2.21-eV photon excitation energy is attributed to a double-resonance enhancement facilitated by direct excitation into a state that contains a small electron polaron.[6]

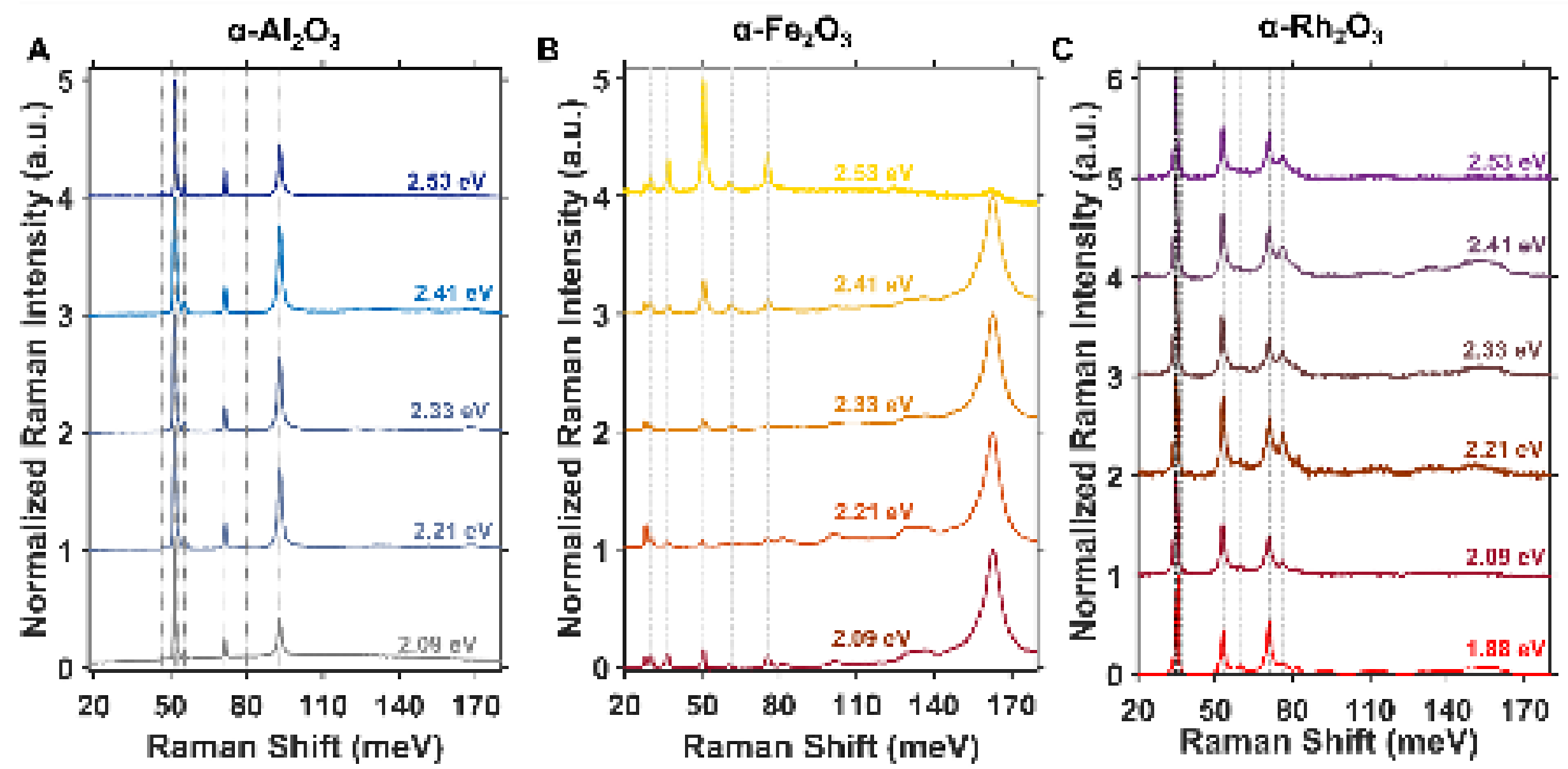

**Figure 5.** Stokes Raman spectra of **(A)** α-$Al_2O_3$, **(B)** α-$Fe_2O_3$, and **(C)** α-$Rh_2O_3$ collected with photon excitation sources ranging from 2.53 – 1.88 eV. Vertical lines are plotted at Raman peak centers of the 7 Raman active modes ($2A_{1g} + 5E_g$); peak center values taken from the 2.53-eV spectrum of for α-$Al_2O_3$, the 2.21 eV-spectrum of α-$Fe_2O_3$, and the 1.88-eV spectrum of α-$Rh_2O_3$. The excitation energy is noted next to each plotted Raman spectrum.

The shifting relative intensities of phonon modes with excitation photon energy in α-$Fe_2O_3$ is evidence of different phonon coupling strengths to different optical transitions; specifically, the most significant phonon enhancement occurs upon excitation at the band-edge (2.21 eV), indicative of a phonon-coupled optical transition. The lack of significant changes to the Raman spectra in α-$Al_2O_3$ upon changing the excitation photon energy further demonstrates that changes observed in α-$Fe_2O_3$ are due to different phonons coupling to different optical transitions: α-$Al_2O_3$ is non-resonant with all photon excitation sources used. Although there are some subtle second order features observed in the resonance Raman spectra of α-$Rh_2O_3$ centered around 160 meV, unlike α-$Fe_2O_3$ we do not observe any drastic changes to the relative intensities of these second-order features in α-$Rh_2O_3$ with changes in excitation photon energy. However, the three closely spaced first-order Raman phonon modes around 35 meV in α-$Rh_2O_3$ do exhibit shifts in relative intensity with changes in excitation photon energy, similar to the behavior of the low-energy phonon modes of α-$Fe_2O_3$ (Fig. 5C, Fig. 6).

Figure 6A plots a closer view of the low-energy modes observed in resonance Raman spectra of α-$Rh_2O_3$ collected with various excitation photon energies. The three low-energy phonon modes of α-$Rh_2O_3$ are close in energy but their relative intensities change with excitation photon energy. For example, when α-$Rh_2O_3$ is excited with 2.53 eV, the most prominent peak is the 34.4-meV phonon with a shoulder at 35.5 meV and a weak shoulder at 36.7 meV. With decreasing excitation photon energy, the shoulder at 36.7 meV remains weak in intensity, but the 35.5-meV phonon mode increases in intensity while the 34.4-meV mode appears as a shoulder. The shifting relative intensities of Raman modes with varied excitation photon energies in α-$Rh_2O_3$ is evidence of different phonon modes exhibiting different magnitudes of coupling with different optical transitions. The intensities of the resonance Raman spectra of α-$Rh_2O_3$ collected at various excitation photon energies were corrected for scattering cross section, and the corrected phonon intensities are overlaid with the imaginary dielectric spectrum in Figure 6C. These data indicate that the entire spectrum exhibits the most significant enhancement upon excitation at 1.88 eV, with the 35.5-meV phonon mode ($E_g$, Figure 6B) exhibiting the strongest enhancement overall. This excitation energy is near the onset of the optical bandgap, suggesting a polaronic state is being directly accessed upon photoexcitation *via* coupling to thermally populated phonon modes. Although the entire Raman spectrum exhibits increased resonance enhancement at 1.88 eV relative to other excitation energies, the shifting relative intensity of the phonon modes with energies around 35 meV indicates that these phonon modes exhibit the strongest coupling to the optical transition.

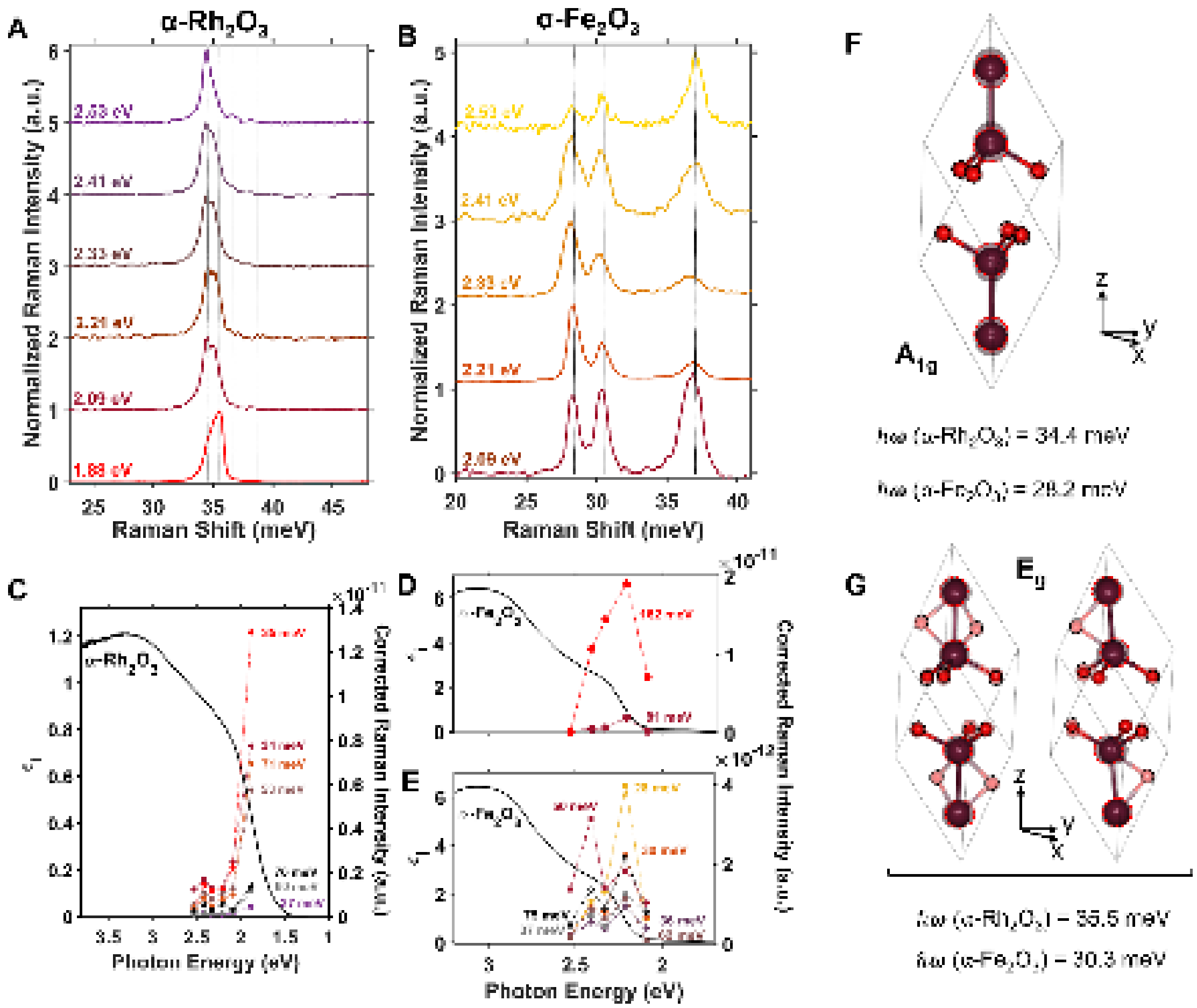

**Figure 6. A, B)** Low frequency Stokes resonance Raman spectra of **(A)** α-$Rh_2O_3$ and **(B)** α-$Fe_2O_3$ collected with photon excitation sources ranging from 2.53 – 1.88 eV. The intensities of all spectra are internally normalized to the most intense peak and vertical lines are plotted to mark the energies of the individual phonon modes. **C-E)** Resonance Raman intensity corrected for scattering cross section of **(C)** first-order phonon modes of α-$Rh_2O_3$, **(D)** the first- and second-order longitudinal optical (LO) modes in α-$Fe_2O_3$, and **(E)** first order phonon modes of α-$Fe_2O_3$. **F, G)** Phonon vector displacement images associated with phonon modes of **(F)** $A_{1g}$ and **(G)** $E_g$ symmetries within the corundum oxide lattice. These modes are labeled with their experimentally determined energies.

The 35.5-meV phonon mode in α-$Rh_2O_3$ (Fig. 6B) matches the symmetry of the 30-meV phonon in α-$Fe_2O_3$ (both are $E_g$). Although the relative intensities of the α-$Fe_2O_3$ Raman spectra shown here do not indicate that the intensity of the 30-meV mode is significantly more enhanced relative to the other phonon modes, first principle calculations considering phonon motions projected across all of k-space indicate that both the 31-meV ($E_g$) and 81-meV (LO) modes of α-$Fe_2O_3$ make significant contributions to the lattice displacement that occurs upon formation of small electron polarons.[7] The lowest-energy phonon mode in α-$Fe_2O_3$ at 28 meV ($A_{1g}$ symmetry) is the first order phonon mode that exhibits the highest intensity when excited with a photon energy of 2.21 eV, which is the energy where the intensity of the 163-meV second-order LO mode peaks and where direct excitation into the polaronic state is observed (Fig. 6C-E).[6] Fourier transformations of transient absorption spectra of α-$Fe_2O_3$ reported by Johnston, *et al.* that were collected with a pump photon energy of 2.25 eV indicate that optical transitions at this pump energy couple strongly to the coherent motion associated with the $A_{1g}$ phonon mode at 28 meV.[48] Molecular dynamics (MD) and time-dependent density functional theory (TDDFT) modeling of photogenerated polaron formation in α-$Fe_2O_3$ further corroborates the strong coupling of these transitions with the $A_{1g}$ phonon mode and suggest that the lattice distortion is associated with formation of a hole polaron.

Figure 7 plots the projected densities of vibrational states of α-$Rh_2O_3$ calculated from Hubbard-corrected DFT. This plot reflects the density of phonon states integrated over all of k-space whereas the Raman spectra only report on the phonon energies at the Γ point. The projected density of vibrational states of α-$Rh_2O_3$ has two well-defined regions: Rh character dominates the vibrational states below 42 meV and oxygen character dominates above 55 meV, with a phonon band gap spanning the energies in between. The displacement vectors corresponding to the Raman-active phonon modes, which represent motion at Γ, corroborate this trend, with majority Rh motion in the lower energy modes and majority oxygen motion in modes with energies greater than 55 meV (Figure 7). Upon careful inspection of the phonon displacement vectors in α-$Rh_2O_3$, it becomes apparent that the three closely spaced phonon modes observed in experimental Raman spectra around 35 meV are all dominated by Rh motion, whereas the higher energy modes (peak center ≥ 55 meV) are dominated by oxygen motion (Figure 7). These trends are analogous to those observed in α-$Fe_2O_3$ where the low energy modes (e.g. at 28 meV and 30 meV) are dominated by Fe motion and the high energy modes (e.g. at 81 meV) are dominated by O motion.[7]

**Figure 7.** Displacement vectors corresponding to the Raman-active phonon modes of α-$Rh_2O_3$ calculated at k-point Γ and aligned in energy with the vibrational density of states evaluated over all k-space. The phonon mode energies shown here were calculated by DFT+*U*.

**Temperature Dependence of Corundum Metal Oxide Optical Transitions**

To further assess the phonon-coupling of optical transitions in corundum metal oxides, thermal difference spectroscopy (TDS) was employed whereby the transmission and reflection spectra were collected at various temperatures and converted to the imaginary dielectric spectrum by applying Fresnel equations.[6,32] Figure 8 shows the thermal difference imaginary dielectric spectra of α-$Rh_2O_3$ and α-$Fe_2O_3$ above (A, C) and (B, D) below room temperature; the intensities of these spectra are defined by Equation 1, where $\epsilon_{i,T}$ represents the magnitude of the imaginary dielectric at a given temperature T.

$$\Delta\epsilon_i(T) = \epsilon_{i,T} - \epsilon_{i,294K} \quad (1)$$

Importantly, in both α-$Rh_2O_3$ and α-$Fe_2O_3$, we observe significant temperature dependence of the optical absorption spectrum in the same energetic region where phonon enhancement is observed in the resonance Raman excitation profiles. Our previous work established that excitation of α-$Fe_2O_3$ at the absorption onset of 2.21 eV directly populates a polaronic excited state, a mechanism that was corroborated by a combination of significant resonance enhancement observed in Raman spectra collected with an excitation photon energy of 2.2 eV and the temperature dependence observed in this region of the dielectric spectrum (Fig. 8C,D).[6,7] In the case of α-$Rh_2O_3$, the resonance Raman excitation profile shows significant enhancement at 1.88 eV, which is about where the derivative shape in the thermal difference spectra peaks (Fig. 8A,B). We note that the derivative shapes of the thermal difference dielectric spectra of α-$Rh_2O_3$ have fewer well-defined peaks compared to α-$Fe_2O_3$, indicating that the transitions in α-$Fe_2O_3$ are more energetically isolated from each other compared to those in α-$Rh_2O_3$.

**Figure 8. A,B)** Thermal difference imaginary dielectric spectra of α-$Rh_2O_3$ **(A)** above room temperature and **(B)** below room temperature. **C)** Integrated and normalized absolute intensities of thermal difference spectra of α-$Rh_2O_3$ plotted versus temperature overlaid with the Bose-Einstein function corresponding to a phonon energy of 35 meV. **D,E)** Analogous TDS plots of α-$Fe_2O_3$ **(D)** above room temperature and **(E)** below room temperature. **F)** Integrated and normalized absolute TDS intensities of α-$Fe_2O_3$ plotted versus temperature and overlaid with a Bose-Einstein function corresponding to a phonon energy of 50 meV. Plots D-F reproduced with permission from reference [6]. Copyright 2021 American Chemical Society.

Figures 8 C,F plot the integrated and normalized absolute values of the intensities of the thermal difference spectra of α-$Rh_2O_3$ and α-$Fe_2O_3$ versus temperature. These plots are normalized to the value of the integrand obtained at 82 K and overlaid with the Bose-Einstein functions corresponding to phonon energies that best match the data. These functions are described by Equations 2 and 3, where N is the phonon population, $\hbar\Omega$ is phonon energy, T is temperature, k is the Boltzmann constant, and $\Delta\epsilon$ is the spectrally integrated magnitude of the thermal difference spectrum.[6]

$$N(\hbar\Omega, T) = (e^{\frac{\hbar\Omega}{kT}} - 1)^{-1} \times \rho(\hbar\Omega) \quad (2)$$

$$\frac{|\Delta\epsilon(T)|}{|\Delta\epsilon(82K)|} = \frac{|N(\hbar\Omega,T)-N(\hbar\Omega,294K)|}{|N(\hbar\Omega,82K)-N(\hbar\Omega,294K)|} \quad (3)$$

For α-$Rh_2O_3$, the phonon energy that generates the Bose-Einstein function that best overlays the TDS data is 35 meV. Due to the similarity in energy of the three lowest energy phonon modes measured with Raman spectroscopy (Fig. 5), combined with the resonance Raman assessment and phonon mode displacement vector assignment (Fig. 6), we assign the phonon modes most strongly coupled to band-edge absorption in α-$Rh_2O_3$ to the 34.4- and 35.5-meV modes. Both of these modes are dominated by Rh-motion and exhibit the largest changes in intensity as the resonance Raman excitation approaches the band-edge onset. Conversely, the threshold phonon energy determined from overlaying the temperature-dependent TDS intensities of α-$Fe_2O_3$ with a Bose-Einstein function is 50 meV, which defines the top edge of the phonon band gap and is dominated by oxygen motion.[7]

To understand the origin of the difference in the nature of the phonon modes that couple most strongly to the band-edge absorption in α-$Rh_2O_3$ and α-$Fe_2O_3$, we compare the electronic and vibrational densities of states of these two materials in Figure 9. Upon close inspection of these plots, two key differences become clear: (i) the Rh 4*d* bands exhibit a higher degree of overlap with O 2*p* bands compared to the Fe 3*d* bands in the electronic structures, suggesting a higher degree of covalency (Fig. 9A), and (ii) the projected density of vibrational states of α-$Rh_2O_3$ has a wider phonon band gap compared to α-$Fe_2O_3$ (Fig. 9B). We assessed the lattice covalency quantitatively using equation 4, in which $\rho_O$ and $\rho_M$ are the projected densities of oxygen and metal states, respectively.[49] Table 2 contains the results of these calculations.

$$c = \frac{\int \rho_O \rho_M dE}{\int \rho_O + \rho_M dE} \quad (4)$$

The calculated degree of covalency of both the valence and conduction band regions of α-$Fe_2O_3$ are significantly lower compared to that calculated for α-$Rh_2O_3$, confirming that α-$Rh_2O_3$ is more covalent than α-$Fe_2O_3$.

**Table 2.** Calculated Degree of Covalency from pDOS of α-$Fe_2O_3$ and α-$Rh_2O_3$

| Band Region | α-$Fe_2O_3$ | α-$Rh_2O_3$ |
|---|---|---|
| Valence Band Edge<br>Fe: -8.85 – 0.5 eV<br>Rh: -9.30 – 0.5 eV | 6.778 | 14.914 |
| Conduction Band Edge<br>Fe: 1.0 – 4.06 eV<br>Rh: 0.75 – 4.43 eV | 1.238 | 3.819 |

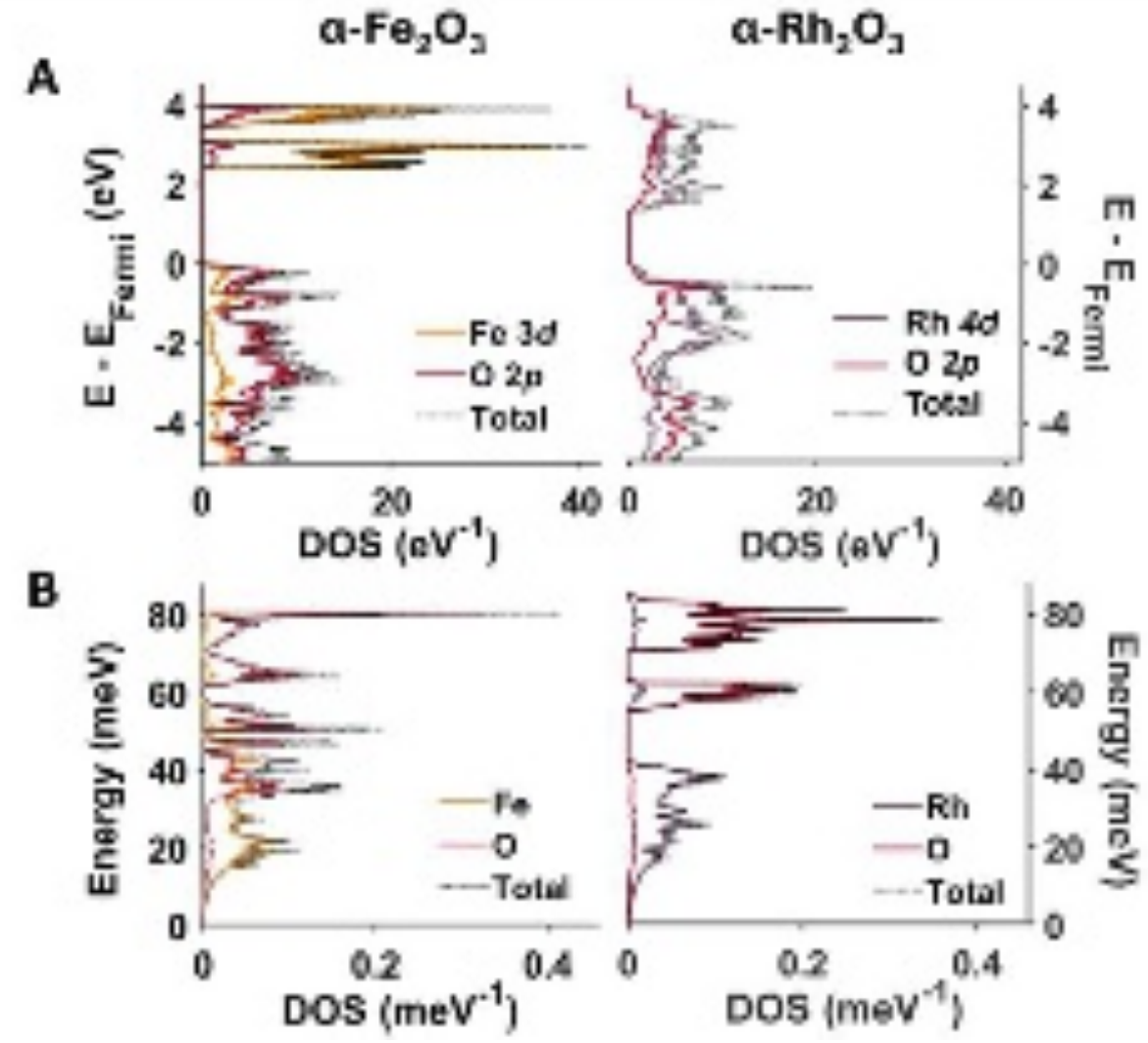

**Figure 9. A)** Electronic projected densities of state of α-$Fe_2O_3$ (*left*) and α-$Rh_2O_3$ (*right*). **B)** Vibrational projected densities of state of α-$Fe_2O_3$ (*left*) and α-$Rh_2O_3$ (*right*). α-$Fe_2O_3$ electronic densities of state reproduced with permission from reference [6]. Copyright 2021 American Chemical Society. Vibrational projected densities of state reproduced with permission from reference [7] with the permission of AIP Publishing.

The projected vibrational densities of state of α-$Rh_2O_3$ has two well-defined regions: Rh character dominating the vibrational states below 42 meV and oxygen character above 55 meV, establishing a phonon band gap (Fig. 9B, *right*). In α-$Fe_2O_3$, there are also Fe-dominated and O-dominated regions present in the projected vibrational densities of state; however, there are regions of overlapping Fe and O contributions between 30 and 50 meV, and the phonon band gap is much narrower in α-$Fe_2O_3$ compared to α-$Rh_2O_3$ (Fig. 9B). We attribute these differences in the projected vibrational densities of states to the fact that there is a much bigger contrast in the atomic masses of Rh and O than for Fe and O. Our analysis of the thermal difference spectra of α-$Fe_2O_3$ indicates that oxygen motion is critical for enabling coupling between thermally populated phonons and optical transitions, as the threshold energy for thermal activation of band-edge absorption coincides with the lowest energy phonon that is dominated by oxygen motion.[6,7] In contrast, the threshold energy for thermal activation of band-edge absorption determined for α-$Rh_2O_3$ corresponds to a phonon mode that is dominated by Rh motion. The higher degree of covalency observed in the electronic structure of α-$Rh_2O_3$ indicates that thermal activation of only phonon modes dominated by Rh is enough to significantly modulate the energies of the valence and conduction bands. Conversely, the ionic nature of the Fe-O bonds in α-$Fe_2O_3$ and the consequently poor overlap of Fe and O electronic states requires enough thermal energy to activate both iron and oxygen-dominated modes in order to enable phonon-coupled optical transitions.

## CONCLUSIONS

A systematic comparison of the optical and Raman spectra of corundum metal oxides reveals that occupied *d* orbitals contribute to the visible absorption of α-$Fe_2O_3$ and α-$Rh_2O_3$. In the absence of occupied *d* orbitals, the band gap of α-$Al_2O_3$ is outside the visible range and, consequently, there is no observed resonance Raman enhancement in this range. The resonance Raman excitation profiles of α-$Fe_2O_3$ and α-$Rh_2O_3$ collected with excitation photon sources spanning the dielectric spectrum of each material demonstrate strong resonance enhancement of specific phonon modes at the onset of absorption in both materials. Thermal difference spectroscopy further corroborates that both α-$Fe_2O_3$ and α-$Rh_2O_3$ exhibit strongly phonon-coupled optical transitions. Quantitative analysis of the resonance Raman excitation profile and thermal difference spectra of α-$Rh_2O_3$ shows that the low-energy phonon modes around 35 meV, which are dominated by Rh motion, exhibit both the strongest resonance enhancement and enable thermal activation of band-edge absorption. In contrast, Bose-Einstein analysis of the TDS of α-$Fe_2O_3$ indicates that thermal population of only Fe-dominated modes is not sufficient to activate band-edge absorption; higher energy O-dominated modes must also be thermally populated. The projected densities of electronic states of α-$Rh_2O_3$ demonstrate a higher degree of covalency between metal and oxygen compared to α-$Fe_2O_3$. The greater covalency in α-$Rh_2O_3$ increases the sensitivity of both the conduction and valence band energies to Rh motion alone and therefore the Rh-dominated phonon modes can couple to band-edge transitions to form photoexcited polarons. Although the data presented cannot definitively differentiate whether small or large polarons form in α-$Rh_2O_3$, these data do demonstrate that increased covalency in a lattice modulates which phonon modes contribute most to the formation of photogenerated polaronic states. Photogenerated polarons have significant impacts on the performance of metal oxide semiconductors in various photoapplications, particularly in photocatalysis.[16,18,19,50] Our results suggest that tuning lattice covalency offers a potential strategy to engineer photoinduced polaron formation pathways to optimize semiconductor performance in these applications.

## SUPPLEMENTARY MATERIAL

The supplementary material contains additional information about computational details along with further optical and structural characterization of α-$Rh_2O_3$ films.

## ACKNOWLEDGMENTS

The authors acknowledge University of Rochester Funding that supported this work and E. P. C. thanks the Elon Huntington Hooker Fellowship for financial support. The authors thank Jacob Shelton for foundational work on the linear response Hubbard and Hund parameter calculations, Fresnel analysis and Raman spectroscopy procedural development. All Raman spectra were collected using the University of Rochester Chemistry Department Multiwavelength Raman Spectroscopy Facility and the authors thank Dave McCamant for continual support on optimizing Raman measurements. All DFT calculations were performed using the University of Rochester Center for Integrated Research and Computing (CIRC); the authors thank the CIRC staff for their support. The authors thank William Brennessel for powder X-ray diffraction collection performed at the University of Rochester X-ray Crystallography Facility.

## AUTHOR DECLARATIONS

### Conflict of Interest Statement

The authors have no conflicts to disclose.

### Author Contributions

E.P.C. made all α-$Rh_2O_3$ samples and collected and processed all original experimental data. E.P.C. performed all original DFT calculations. E.P.C. wrote the manuscript. Both authors contributed to editing the manuscript. K.E.K. supervised the project.

**DATA AVAILABILITY STATEMENT**

The data that support the findings of this study are available from the corresponding author upon reasonable request.

**REFERENCES**

[1] A. Imtiyaz, and A. Singh, "Transition metal oxides (TMOs) as Photocatalysts: an updated review of synthesis, applications, and future implications," Graphene and 2D Mater **9**(3–4), 187–226 (2024).

[2] A. Krishnan, A. Swarnalal, D. Das, M. Krishnan, V.S. Saji, and S.M.A. Shibli, "A review on transition metal oxides based photocatalysts for degradation of synthetic organic pollutants," Journal of Environmental Sciences **139**, 389–417 (2024).

[3] X. Yu, T.J. Marks, and A. Facchetti, "Metal oxides for optoelectronic applications," Nature Mater **15**(4), 383–396 (2016).

[4] S.Mohd. Waseem, S.V. Gaikwad, V.A. Pandit, and N.N. Kapse, "Iron Oxide Nanoparticles for Catalytic Applications: Synthesis, Characterization, and Environmental Performance," Top Catal, (2025).

[5] S. Lany, "Semiconducting transition metal oxides," J. Phys.: Condens. Matter **27**(28), 283203 (2015).

[6] J.L. Shelton, and K.E. Knowles, "Thermally Activated Optical Absorption into Polaronic States in Hematite," J. Phys. Chem. Lett. **12**(13), 3343–3351 (2021).

[7] J.L. Shelton, and K.E. Knowles, "Polaronic optical transitions in hematite (α-Fe2O3) revealed by first-principles electron–phonon coupling," The Journal of Chemical Physics **157**(17), 174703 (2022).

[8] L.D. Landau, "Electron Motion in Crystal Lattices," Phys. Z. Sowjetunion **3**, 664 (1933).

[9] L.D. Landau, and Pekar, "Effective mass of polaron," Zh. Eksp. Teor. Fiz. **18**, 419 (1948).

[10] C. Franchini, M. Reticcioli, M. Setvin, and U. Diebold, "Polarons in materials," Nat Rev Mater **6**(7), 560–586 (2021).

[11] D.A. Grave, N. Yatom, D.S. Ellis, M.C. Toroker, and A. Rothschild, "The 'Rust' Challenge: On the Correlations between Electronic Structure, Excited State Dynamics, and Photoelectrochemical Performance of Hematite Photoanodes for Solar Water Splitting," Advanced Materials **30**(41), 1706577 (2018).

[12] Y. Lin, G. Yuan, S. Sheehan, S. Zhou, and D. Wang, "Hematite-based solar water splitting: challenges and opportunities," Energy Environ. Sci. **4**(12), 4862 (2011).

[13] S. Shen, S.A. Lindley, X. Chen, and J.Z. Zhang, "Hematite heterostructures for photoelectrochemical water splitting: rational materials design and charge carrier dynamics," Energy Environ. Sci. **9**(9), 2744–2775 (2016).

[14] K. Sivula, F. Le Formal, and M. Grätzel, "Solar Water Splitting: Progress Using Hematite (α-$Fe_2 O_3$ ) Photoelectrodes," ChemSusChem **4**(4), 432–449 (2011).

[15] C. Di Valentin, and A. Selloni, "Bulk and Surface Polarons in Photoexcited Anatase TiO2," J. Phys. Chem. Lett. **2**(17), 2223–2228 (2011).

[16] M. Reticcioli, I. Sokolović, M. Schmid, U. Diebold, M. Setvin, and C. Franchini, "Interplay between Adsorbates and Polarons: CO on Rutile TiO 2 ( 110 )," Phys. Rev. Lett. **122**(1), 016805 (2019).

[17] Y. Cao, M. Yu, S. Qi, S. Huang, T. Wang, M. Xu, S. Hu, and S. Yan, "Scenarios of polaron-involved molecular adsorption on reduced TiO2(110) surfaces," Sci Rep **7**(1), 6148 (2017).

[18] S. Bandaranayake, A. Patnaik, E. Hruska, Q. Zhu, S. Das, and L.R. Baker, "Effect of Surface Electron Trapping and Small Polaron Formation on the Photocatalytic Efficiency of Copper(I) and Copper(II) Oxides," ACS Appl. Mater. Interfaces **16**(31), 41616–41625 (2024).

[19] H. Gajapathy, S. Bandaranayake, E. Hruska, A. Vadakkayil, B.P. Bloom, S. Londo, J. McClellan, J. Guo, D. Russell, F.M.F. Groot, F. Yang, D.H. Waldeck, M. Schultze, and L.R. Baker, "Spin polarized electron dynamics enhance water splitting efficiency by yttrium iron garnet photoanodes: a new platform for spin selective photocatalysis," Chem. Sci. **15**(9), 3300–3310 (2024).

[20] H. Fröhlich, "Electrons in lattice fields," Advances in Physics **3**(11), 325–361 (1954).

[21] H. Fröhlich, H. Pelzer, and S. Zienau, "XX. Properties of slow electrons in polar materials," The London, Edinburgh, and Dublin Philosophical Magazine and Journal of Science **41**(314), 221–242 (1950).

[22] T. Holstein, "Studies of polaron motion: Part I. The molecular-crystal model," Annals of Physics **8**(3), 325–342 (1959).

[23] T. Holstein, "Studies of polaron motion: Part II. The 'small' polaron," Annals of Physics **8**(3), 343–389 (1959).

[24] M. Reticcioli, U. Diebold, G. Kresse, and C. Franchini, “Small Polarons in Transition Metal Oxides,” in *Handbook of Materials Modeling*, edited by W. Andreoni and S. Yip, (Springer International Publishing, Cham, 2020), pp. 1035–1073.

[25] H. Hassani, E. Bousquet, X. He, B. Partoens, and P. Ghosez, “The anti-distortive polaron as an alternative mechanism for lattice-mediated charge trapping,” Nat Commun **16**(1), 1688 (2025).

[26] L.M. Carneiro, S.K. Cushing, C. Liu, Y. Su, P. Yang, A.P. Alivisatos, and S.R. Leone, “Excitation-wavelength-dependent small polaron trapping of photoexcited carriers in α-Fe2O3,” Nature Mater **16**(8), 819–825 (2017).

[27] J. Husek, A. Cirri, S. Biswas, and L.R. Baker, “Surface electron dynamics in hematite (α-$Fe_2O_3$): correlation between ultrafast surface electron trapping and small polaron formation,” Chem. Sci. **8**(12), 8170–8178 (2017).

[28] E.P. Craddock, J.L. Shelton, M.T. Ruggiero, and K.E. Knowles, “Local coordination geometry within cobalt spinel oxides mediates photoinduced polaron formation,” Chem. Sci. **16**(24), 10759–10770 (2025).

[29] S. Restelli, O. Cannelli, N. Colonna, C. Grova, P. Usai, M. Puppin, M. Mensi, F. Barantani, Y. Meng, J. Teyssier, M. Oppermann, F. Pennacchio, C. Bacellar, J.R. Rouxel, L.D. Leroy, O. Dogadov, N. Ohannessian, D. Pergolesi, P. Galinetto, P. Piekarz, A. Ptok, S. Chaudhary, G.A. Fiete, M. Chergui, E. Baldini, and G.F. Mancini, “Ultrafast Formation of Jahn–Teller Polarons Revealed by State-Selective Excitation in Correlated Spinel $Co_3O_4$,” J. Am. Chem. Soc., jacs.5c23346 (2026).

[30] H. Liu, H. Cong, G. Yang, C. Gong, J. Ding, Y.Y. Fu, J. Cui, K. Song, B. Chen, C. He, N. Zhao, J. Ye, and F. He, “Surface hole polaron site tuning governs charge carrier separation in BiVO4 photoanodes,” Nat Commun **17**(1), 2562 (2026).

[31] H. Wu, L. Zhang, S. Qu, A. Du, J. Tang, and Y.H. Ng, “Polaron-Mediated Transport in $BiVO_4$ Photoanodes for Solar Water Oxidation,” ACS Energy Lett. **8**(5), 2177–2184 (2023).

[32] A. Barybin, and V. Shapovalov, “Substrate Effect on the Optical Reflectance and Transmittance of Thin-Film Structures,” International Journal of Optics **2010**, 1–18 (2010).

[33] M. Cococcioni, and S. De Gironcoli, “Linear response approach to the calculation of the effective interaction parameters in the LDA + U method,” Phys. Rev. B **71**(3), 035105 (2005).

[34] B. Himmetoglu, R.M. Wentzcovitch, and M. Cococcioni, “First-principles study of electronic and structural properties of CuO,” Phys. Rev. B **84**(11), 115108 (2011).

[35] B. Himmetoglu, A. Floris, S. De Gironcoli, and M. Cococcioni, "Hubbard-corrected DFT energy functionals: The LDA+U description of correlated systems," Int. J. Quantum Chem. **114**(1), 14–49 (2014).

[36] O. Neufeld, and M. Caspary Toroker, "Play the heavy: An effective mass study for α-Fe2O3 and corundum oxides," The Journal of Chemical Physics **144**(16), 164704 (2016).

[37] N. Hiraoka, H. Okamura, H. Ishii, I. Jarrige, K.D. Tsuei, and Y.Q. Cai, "Charge transfer and dd excitations in transition metal oxides: q-dependent dielectric function study by inelastic X-ray scattering and optical reflectivity measurements," Eur. Phys. J. B **70**(2), 157–162 (2009).

[38] M. Sachs, L. Harnett-Caulfield, E. Pastor, B. Davies, D.J.C. Sowood, B. Moss, A. Kafizas, J. Nelson, A. Walsh, and J.R. Durrant, "Metal-centred states control carrier lifetimes in transition metal oxide photocatalysts," Nat. Chem. **17**(9), 1348–1355 (2025).

[39] R. Rajan, E.P. Craddock, and K.E. Knowles, "Impact of Composition on the Optical Properties of Colloidal Ternary Spinel Oxide Nanocrystals: Spinel Ferrites versus Spinel Gallates," Chem. Mater. **38**(9), 4561–4569 (2026).

[40] I.R. Beattie, and T.R. Gilson, "The single-crystal Raman spectra of nearly opaque materials. Iron(III) oxide and chromium(III) oxide," J. Chem. Soc., A, 980 (1970).

[41] S.P.S. Porto, and R.S. Krishnan, "Raman Effect of Corundum," The Journal of Chemical Physics **47**(3), 1009–1012 (1967).

[42] W.H. Weber, R.J. Baird, and G.W. Graham, "Raman investigation of palladium oxide, rhodium sesquioxide and palladium rhodium dioxide," J Raman Spectroscopy **19**(4), 239–244 (1988).

[43] C.P. Marshall, W.J.B. Dufresne, and C.J. Rufledt, "Polarized Raman spectra of hematite and assignment of external modes," J Raman Spectroscopy **51**(9), 1522–1529 (2020).

[44] J.M.D. Coey, "The crystal structure of $Rh_2O_3$," Acta Crystallogr B Struct Sci **26**(11), 1876–1877 (1970).

[45] C.P. Marshall, and W.J.B. Dufresne, "Resonance Raman and polarized Raman scattering of single-crystal hematite," J Raman Spectroscopy **53**(5), 947–955 (2022).

[46] K. Nakamoto, *Infrared and Raman Spectra of Inorganic and Coordination Compounds: Part A: Theory and Applications in Inorganic Chemistry*, 1st ed. (Wiley, 2008).

[47] *Introductory Raman Spectroscopy* (Elsevier, 2003).

[48] C.R. Johnston, R. Speelman, A. Arcidiacono, C. Bridgewater, L. Rassouli, X. Ma, I. Vargas-Hurlston, J. Kupferberg, L. Martin, A.B.F. Martinson, M. Dupuis, F.M. Geiger, and S.B. King,

"Watching Polarons Dance: Coherent Carrier–Phonon Coupling in Hematite Revealed by Transient Absorption Spectroscopy," J. Am. Chem. Soc. **147**(44), 40338–40346 (2025).

[49] P.C. Müller, C. Ertural, J. Hempelmann, and R. Dronskowski, "Crystal Orbital Bond Index: Covalent Bond Orders in Solids," J. Phys. Chem. C **125**(14), 7959–7970 (2021).

[50] E.P. Craddock, and K.E. Knowles, "Signatures of photogenerated small polarons in transition metal oxides," Chem. Commun. **62**(10), 3151–3166 (2026).

[51] P. Giannozzi, S. Baroni, N. Bonini, M. Calandra, R. Car, C. Cavazzoni, D. Ceresoli, G.L. Chiarotti, M. Cococcioni, I. Dabo, A. Dal Corso, S. De Gironcoli, S. Fabris, G. Fratesi, R. Gebauer, U. Gerstmann, C. Gougoussis, A. Kokalj, M. Lazzeri, L. Martin-Samos, N. Marzari, F. Mauri, R. Mazzarello, S. Paolini, A. Pasquarello, L. Paulatto, C. Sbraccia, S. Scandolo, G. Sclauzero, A.P. Seitsonen, A. Smogunov, P. Umari, and R.M. Wentzcovitch, "QUANTUM ESPRESSO: a modular and open-source software project for quantum simulations of materials," J. Phys.: Condens. Matter **21**(39), 395502 (2009).

[52] P. Giannozzi, O. Andreussi, T. Brumme, O. Bunau, M. Buongiorno Nardelli, M. Calandra, R. Car, C. Cavazzoni, D. Ceresoli, M. Cococcioni, N. Colonna, I. Carnimeo, A. Dal Corso, S. De Gironcoli, P. Delugas, R.A. DiStasio, A. Ferretti, A. Floris, G. Fratesi, G. Fugallo, R. Gebauer, U. Gerstmann, F. Giustino, T. Gorni, J. Jia, M. Kawamura, H.-Y. Ko, A. Kokalj, E. Küçükbenli, M. Lazzeri, M. Marsili, N. Marzari, F. Mauri, N.L. Nguyen, H.-V. Nguyen, A. Otero-de-la-Roza, L. Paulatto, S. Poncé, D. Rocca, R. Sabatini, B. Santra, M. Schlipf, A.P. Seitsonen, A. Smogunov, I. Timrov, T. Thonhauser, P. Umari, N. Vast, X. Wu, and S. Baroni, "Advanced capabilities for materials modelling with Quantum ESPRESSO," J. Phys.: Condens. Matter **29**(46), 465901 (2017).

[53] P. Giannozzi, O. Baseggio, P. Bonfà, D. Brunato, R. Car, I. Carnimeo, C. Cavazzoni, S. De Gironcoli, P. Delugas, F. Ferrari Ruffino, A. Ferretti, N. Marzari, I. Timrov, A. Urru, and S. Baroni, "Q UANTUM ESPRESSO toward the exascale," The Journal of Chemical Physics **152**(15), 154105 (2020).

[54] J.P. Perdew, K. Burke, and M. Ernzerhof, "Generalized Gradient Approximation Made Simple," Phys. Rev. Lett. **77**(18), 3865–3868 (1996).

[55] J.P. Perdew, A. Ruzsinszky, G.I. Csonka, O.A. Vydrov, G.E. Scuseria, L.A. Constantin, X. Zhou, and K. Burke, "Restoring the Density-Gradient Expansion for Exchange in Solids and Surfaces," Phys. Rev. Lett. **100**(13), 136406 (2008).

[56] D.R. Hamann, "Optimized norm-conserving Vanderbilt pseudopotentials," Phys. Rev. B **88**(8), 085117 (2013).

[57] M.J. Van Setten, M. Giantomassi, E. Bousquet, M.J. Verstraete, D.R. Hamann, X. Gonze, and G.-M. Rignanese, "The PseudoDojo: Training and grading a 85 element optimized norm-conserving pseudopotential table," Computer Physics Communications **226**, 39–54 (2018).

[58] R. Fletcher, editor , *Practical Methods of Optimization*, 2nd ed (Wiley, Chichester New York, 2010).

[59] S.R. Billeter, A.J. Turner, and W. Thiel, "Linear scaling geometry optimisation and transition state search in hybrid delocalised internal coordinates," Phys. Chem. Chem. Phys. **2**(10), 2177–2186 (2000).

[60] S.R. Billeter, A. Curioni, and W. Andreoni, "Efficient linear scaling geometry optimization and transition-state search for direct wavefunction optimization schemes in density functional theory using a plane-wave basis," Computational Materials Science **27**(4), 437–445 (2003).